\documentclass{IEEEtran}
\usepackage{cite}
\usepackage{amsmath,amssymb,amsfonts}
\usepackage{algorithmic}
\usepackage{graphicx}
\usepackage{textcomp}
\def\BibTeX{{\rm B\kern-.05em{\sc i\kern-.025em b}\kern-.08em
    T\kern-.1667em\lower.7ex\hbox{E}\kern-.125emX}}
\usepackage{siunitx}
\usepackage{hyperref}
\usepackage{xcolor}
\usepackage{soulutf8}
\usepackage{enumitem}

\begin{document}
\title{Articulatory movements influence electromagnetic wave transmission through the vocal tract}
\author{Rémi Blandin, Martin Laabs, Rudolf von Bünau, Bryn Lloyd, Silvia Farcito, Denys Nikolayev, Gabriela Hossu, Peter~Birkholz, and Dirk Plettemeier,
\thanks{This study was funded by the German Federal Ministry of Educationand  Research  (BMBF)
within  the  project  “Promise-AI,”  Funding  Code No.  16SV8988/9,  which  was  a  joint
project  between  TU  Dresden  and Altavo GmbH. 
The MRI data collection project conducted in Nancy (study/ work) is co-funded by the French State-Region contract CPER 2015-2020 (Contrat de Plan Etat Région– IT2MP Innovations Technologiques, Modélisation et Médecine Personnalisée) and by the European Union through the European Regional Development Fund “FEDER-FSE Lorraine et Massif des Vosges 2014-2020”.
This work was performed on a platform member of France Life Imaging network (grant ANR-11-INBS-0006).
}
\thanks{R. Blandin, was with 
Institute of Acoustics and Speech Communication,
Technische Universität Dresden (TU Dresden), 01069 Dresden, Germany. He is now with 
Université de Rennes, CNRS, IETR-UMR 6164; F-35000 Rennes, France (e-mail: remi.blandin@univ-rennes.fr).}
\thanks{M. Laabs is with Institute of Communication Technologies (iFN),
Technische Universität Dresden (TU Dresden), 01069 Dresden, Germany (e-mail: martin.laabs@tu-dresden.de).}
\thanks{R. von Bünau is with 
Altavo GmbH 01187 Dresden, Germany (e-mail: rudolf.vonbuenau@altavo.eu).}
\thanks{B. Lloyd is with Virtual Population Group
Foundation for Research on Information Technologies in Society (IT'IS)
8004 Zurich, Switzerland (e-mail: lloyd@itis.swiss).}
\thanks{S. Farcito is with Virtual Population Group
Foundation for Research on Information Technologies in Society (IT'IS)
8004 Zurich, Switzerland (e-mail: farcito@itis.swiss).}
\thanks{D. Nikolayev is with Université de Rennes, CNRS, IETR-UMR 6164; F-35000 Rennes, France 
 (e-mail: d@deniq.com).}
\thanks{G. Hossu is with Université de Lorraine, Inserm, IADI, F-54000 Nancy, France
CHRU-Nancy, INSERM, Université de Lorraine, CIC, Innovation Technologique, F-54000 Nancy, France.}
\thanks{P. Birkholz is with Institute of Acoustics and Speech Communication,
Technische Universität Dresden (TU Dresden),
01069 Dresden, Germany (e-mail: peter.birkholz@tu-dresden.de).}
\thanks{D. Plettemeier is with Institute of Communication Technologies (iFN),
Technische Universität Dresden (TU Dresden), 01069 Dresden, Germany (e-mail: dirk.plettemeier@tu-dresden.de).}
}

\maketitle

\begin{abstract}
This study experimentally validates a numerical model of electromagnetic propagation through the human head during the pronunciation of different vowels, with the goal of improving our understanding of the underlying physical phenomena.
A realistic finite element model was created from magnetic resonance images acquired while pronouncing the vowels /a/, /i/, and /u/. 
The model was validated against scattering matrix measurements obtained from two subjects whose geometries were modeled. 
Despite several potential sources of discrepancy, the simulations and measurements showed good qualitative agreement, confirming the validity of the approach. 
Similar transmission coefficient patterns were observed across subjects for the same vowels.
Within the investigated frequency range of (1-6~GHz), the electric field exhibited a Mie scattering pattern. 
Local minima and maxima in the transmission coefficient, characterizing different articulatory configurations, were correlated with local variations in the electric field amplitude. 
The transmission coefficient's shape results from an interplay between resonance patterns and antenna placement, while the degree of mouth opening influences the shape of scattering modes.
Although technically challenging, this numerical approach proved effective for studying electromagnetic propagation in the human head. 
The resulting robust numerical model and improved understanding of the underlying physics are expected to facilitate the development of radio-frequency-based silent speech interfaces.
\end{abstract}


\section{Introduction}

\begin{figure*}
    \centering
    \includegraphics[scale=1.25]{figure_1.pdf}
    \caption{
    a) Two antennas placed on either sides of the mouth are used to measure the transmission spectrum, b) the articulatory movements change the electromagnetic field resonance patterns, c) this affects the transmission spectrum, d) which can be used to train a neural network to recognize speech and generate a substitute voice. 
    }
    \label{fig:radar_speech_concept}
\end{figure*}

Radio frequencies (RF) have proven to be an efficient tool for investigating body structure and movements, as demonstrated by a review on the use of RF to observe the human body \cite{gu2014} and another one more specifically targeting vital signs detection \cite{kebe2020human}, such as heartbeat and respiration  \cite{Hahnel2013}\cite{Jang2021}. Beyond these applications, RF technology is also utilized in imaging \cite{Jacobi1986, Thiel2009}, particularly for breast tissues to detect cancer \cite{Moll2014}.

One notable application within this domain is speech movement tracking, which is the focus of the present study. The concept of using an RF system to sense articulation, often referred to as radar speech, was first introduced by Holzrichter~\emph{et~al.} \cite{Holzrichter1998, Holzrichter2009}.
The application of this concept to the generation of a substitute voice is represented in Fig.~\ref{fig:radar_speech_concept}.

This innovative approach presents a promising solution for silent speech interfaces (SSI), which aim to recover speech in the absence of the acoustic signal. SSI systems utilize coexistent, non-audible (bio)signals that are generated during speech production, even when the acoustic signal is absent. These technologies rely on a wide variety of methods, as illustrated in several reviews \cite{Denby2010, gonzalez_2020}.

SSI serve a multitude of purposes, including voice restoration for laryngectomy patients \cite{Fagan2007, Meltzner2017}, enabling private conversations in public areas, and facilitating communication in noisy environments \cite{Denby2010}. Despite their potential, most current techniques are either invasive or impractical for real-world applications. Consequently, practical applications outside of research settings remain limited and largely anecdotal.

The newly developing radar speech method offers a promising technique to overcome these limitations. It is one of the most suitable approaches for practical applications due to its non-invasive nature and reliability, making it a viable option for future advancements in SSI based technologies.

Most studies on radar speech have concentrated on contactless technologies that utilize antennas positioned at a distance from the speaker  \cite{Lee2023, Zhang2024, Steinmetz2024, Li2023}. However, these methods primarily capture external changes in facial geometry, which can often be detected with better signal quality, for example,  using optical depth cameras \cite{Wang2024}.
In fact, a study on contactless technologies for measuring heart rate and respiration demonstrated that such technologies predominantly detect movements on the body surface \cite{aardal2012physical}. This limitation arises from the significant impedance mismatch between air and body tissues, causing most waves to reflect at the interface.
Nevertheless, capturing information about internal articulatory movements, such as those of the tongue, is crucial for achieving high-quality speech recognition. On the other hand, distant antennas are impractical for applications that focus more on wearable devices.

To address the limitation of impedance mismatch, some studies have explored placing antennas in direct contact with the subject's skin \cite{Birkholz2018, Wagner2022}, as illustrated in Figs.~\ref{fig:radar_speech_concept}a and \ref{fig:exp_setup}. This approach enhances the penetration of electromagnetic waves into the head, allowing for the capture of information regarding internal structures, such as the tongue.
Articulatory movements alter the resonance patterns of the electromagnetic field, as shown in Fig.~\ref{fig:radar_speech_concept}b, which in turn affects the transmission spectrum, as illustrated in Fig.~\ref{fig:radar_speech_concept}c. These changes in the transmission spectrum can be utilized to train a neural network. The network can then recognize speech and generate a synthetic substitute voice, as demonstrated in Fig.~\ref{fig:radar_speech_concept}d.
While there are relatively few studies on contact radar speech, the results are promising. For instance, one study achieved a recognition accuracy of 99.17\% for a command-word recognition task conducted across multiple sessions with a single speaker \cite{Wagner2022}. 
An accuracy of 88.87\% was achieved when using sessions from two different speakers.
In a subsequent study involving eight subjects, similar performance was reported --up to 97.78\% for a single subject and 88.94\% across multiple subjects \cite{Menezes2025}.
Another investigation focused on continuous speech recognition for real time application, employing phoneme recognition. This approach yielded a phoneme error rate of 45.62\% which was 15\% higher than that obtained with a benchmark acoustic feature set \cite{Menezes2025_2}.
Additionally, one study examined the combination of radar signals with optical palatography to enhance the robustness of command word recognition tasks. 
The radar-based system alone achieved an accuracy of 85.40\%, compared to 66.99\% for optical palatography alone, while combining both modalities resulted in a slightly higher accuracy of 86.55\% \cite{Menezes2025_3}.
This technology is also the foundation of a commercial application under development:
the Promise-AI project, which funded this work, aims to explore the technological basis for potential speech prostheses using this technology.
Therefore, our focus is on this particular type of radar speech technology.

Before this study, the physical phenomena underlying this technology were poorly understood
and optimization efforts were largely conducted through trial and error without a deep comprehension of the underlying processes \cite{Wagner2022, Possamai2022, Menezes2025}. Furthermore, there was no simulation of the physical phenomena involved, which would enable reproducible testing of the technology.

Therefore, the goal of this study is to develop and experimentally validate an accurate simulation framework for the physical phenomena associated with radar speech. This framework will enhance our understanding of the operational principles of this emerging technology and facilitate its development, for example by enabling prototype design testing and optimizing antenna placement.

Numerical simulations have been extensively employed to understand and optimize RF technologies used on or near the body. They are particularly valuable for evaluating the user exposure and the specific absorption rate (SAR) \cite{Parsa2023, Scarella2006, Simunic1996, Kalloch2019}. Additionally, these simulations are applied in more specialized contexts, such as evaluating a concept for the early diagnosis of Alzheimer's disease \cite{Elloian2014} and detecting brain strokes using an antenna array \cite{Qureshi2018}.

While finite difference methods have been widely used due to their straightforward grid definition from medical images \cite{Parsa2023}, some studies have favored numerical methods that rely on unstructured grids, such as the finite elements method (FEM). These methods offer a more accurate description of anatomical details \cite{Scarella2006, Simunic1996, Kalloch2019}. Given the high accuracy requirements for studying speech, FEM was a relevant choice for our objectives. Consequently, we selected the Ansys High-Frequency Structure Simulator (HFSS)\cite{hfss} as our simulation tool.

Bow-tie dipole antennas are appreciated for their broadband capabilities, simple and easy-to-fabricate design, stable directivity pattern, and consistent polarization. These features, combined with their geometric suitability for attachment on the body, make them a common choice for biomedical applications \cite{Jalilvand2014, Arayeshnia2020, Fiser2022}. For these reasons, we selected this type of antenna design for our measurements.

We created detailed and accurate three-dimensional (3D) models of human heads based on magnetic resonance imaging (MRI) for three distinct vowels: /a/, /i/, and /u/. These vowels represent the most extreme vocal tract positions, thereby covering the full range of variability. To achieve this, we acquired MRI scans of two subjects pronouncing these vowels.

The MRI scans were segmented to identify various tissues, including gray matter, white matter, cerebrospinal fluid, internal air, skull, mandible, teeth, and eyes. The resulting segmentation masks were converted into 3D surface meshes, which were importable into HFSS using a custom-made program. From these 3D models, we constructed numerical FEM models.

We simulated the propagation of electromagnetic waves through these numerical models and validated the simulations by measuring the scattering matrix with a vector network analyzer (VNA) on the same subjects. Bow-tie antennas were attached to each cheek, positioned near the lips, as this location has been shown to be favorable \cite{Wagner2022, Possamai2022, Menezes2025}. The scattering matrices obtained from the measurements were compared with the simulated data, and the electric field was analyzed to investigate the underlying phenomena.

\section{Methods}

\subsection{Ethical approval}

The study received ethical approval from an ethics committee (“EDEN” study; ClinicalTrials.gov Identifier: NCT05218460, approval: CPP SUD-EST IV, 26.07.21). Oral and written informed consent was obtained from each participant before inclusion.
The sponsor was CHRU de Nancy (Département Méthodologie, Promotion, Investigation). This study was supported by the CIC-IT of Nancy.

\subsection{MRI Data Collection}

MRI data were acquired at a different institution for each subject, as the second subject was added in a subsequent phase of the study to expand the dataset.

For Subject~1 it was done
in the Max Planck Institute for Multidisciplinary Sciences Biomedical NMR research group in Göttingen
\cite{MaxPlanckInst}
using the method described in \cite{voit2020} on a 3T system (Magnetom Prisma fit; Siemens Healthineers, Erlangen, Germany)  equipped with a 64-channel head coil. The sequence employed was a radiofrequency (RF)-spoiled gradient echo MRI (FLASH) with mild T1 weighting, high-speed imaging with significant undersampling and automatic slice shifting. The slices were oriented in the sagittal plane, with 9 radial spokes utilized. The imaging parameters included a repetition time of 3.33 ms, an echo time of 2.11 ms, an RF flip angle of 10°, and an acquisition time of 30 ms. The slices had a thickness of 2 mm with a 50\% slice shift, achieving a resolution of 1.5 mm × 1.5 mm × 1 mm. 
The field of view measured 256 × 256 mm², with a matrix size of 170 × 170. A total of 120 images/slices were acquired, covering a lateral width of 120 mm. The total acquisition time was 120 × 30 ms, resulting in a duration of 3.6~s.

Two overlapping images were acquired for each vocal tract configuration. The first acquisition focused on one hemisphere, with the subject maintaining a stable articulatory position for at least 3.6 seconds. The same process was repeated for the opposite hemisphere.

The MRI of Subject~2 was aquired at the CHRU of Nancy, in France using a 3T system (Siemens Prisma; Siemens, Erlangen, Germany) equipped with a 64-channel head coil.
The sequence used was a 3D gradient echo with partial Fourier and parallel imaging (3D VIBE). 
The imaging parameters included a repetition time of 3.57~ms, an echo time of 1.33~ms, an RF flip angle of 9°, an acquisition time of 15~s, a pixel bandwidth of 444~Hz$/$pixel, an acceleration factor iPAT~$=$~3, reference lines 24~AP~×~24~RL, a partial Fourier phase of 6$/$8 and a partial Fourier slice of 7$/$8. 
The slices had a thickness of 1.2~mm achieving a resolution of 1.05469~mm × 1.05469~mm × 1.2~mm in a field of view of 253 ×~ 270 × 144 $\text{mm}^3$. 
A single image was acquired for each vocal tract configuration.

Subject~1 was a 54-year-old male native German speaker, and Subject~2 was a 33-year-old male native French speaker.
The subjects began articulating the target phonemes /a, i, u/ just before each acquisition and maintained as stable an articulatory position as possible throughout the entire duration. Additionally, a reference image was taken with the mouth in a neutral position: lips closed and the tongue relaxed. This process was designed to minimize disturbances from involuntary movements by keeping the acquisition time as short as possible.

\subsection{MRI Segmentation}

To generate phoneme-specific finite element models for Subject~1 we employed a structured image segmentation workflow to minimize inconsistencies between models. First, high-resolution T1 and T2 MR images in a neutral mouth position were manually segmented to create a template model, which included anatomical structures such as gray matter, white matter, cerebrospinal fluid, internal air, skull, mandible, teeth (upper and lower), eyes, and a general head mask. Manual segmentation was performed using iSEG \cite{iSEG}.

The workflow to segment fast MRI scans acquired during phoneme articulation was implemented in Python using SimpleITK \cite{lowekamp2013}. The first step was to pre-process the fast MRI data, including merging of left and right hemisphere scans, resampling to 1x1x1 mm resolution, and N4 bias correction \cite{tustison2010n4itk}. Next, a head mask was segmented using Otsu thresholding \cite{xu2011} and morphological operations. The template model served as the foundation for segmenting fast MRI scans acquired during phoneme articulation, ensuring that differences between models reflected only mouth movements. Since the registration was intra-subject, rigid alignment of bones in the template model to the phoneme-specific images provided a reliable method for transferring anatomical structures. First, the skull was registered to the phoneme images using a rigid transformation, allowing the transfer of tissues enclosed or attached to the skull (teeth, brain and eyes). The mandible (and lower teeth) was then registered separately to account for potential movement relative to the skull during phoneme articulation. The registration approach used Mattes mutual information as a similarity metric \cite{mattes2001}.

The remaining tissues critical to articulation---such as internal air, mucosa, tongue, and velum---were then segmented manually in iSEG. Skin was added with a uniform thickness of 2~mm. All segmentations were reviewed by a medical professional for accuracy and corrected as needed. 

Because a significant time interval separated the processing of Subject~1 and Subject~2, Subject~2 benefited from a more advanced approach unavailable during the processing of Subject~1.
Subject~2 was segmented automatically using an AI-based approach in Sim4Life that directly segmented 40 tissues from the T1 image using a UNet model described in
\cite{Karimi2025}.
Since the automatic segmentation did not perform that well in the vicinity of the teeth (teeth and air are black in MRI), we manually corrected this region in iSEG. Similar to Subject~1, we registered one bone template mask (from phoneme /a/) to the other phonemes to ensure they have consistent skull/mandible geometry and thickness. Finally, we manually segmented the velum, refined the segmentation of the air and tongue, and mapped the 40 tissue labels to the corresponding labels in Subject~1 to have consistent/comparable material assignments.

\subsection{Antenna}

\begin{figure}
    \centering
    \includegraphics[width=\linewidth, trim=7mm 5mm 7mm 2mm, clip]{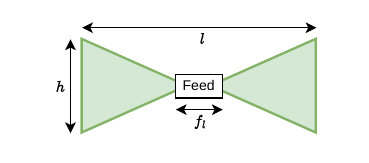}
    \caption{Conductor geometry of the utilized bow-tie antenna, manufactured on \SI{0.8}{\milli\metre} FR4 PCB with dimensions $l = \SI{27}{\milli\metre}$, $h = \SI{11.5}{\milli\metre}$ and $f_l = \SI{3}{\milli\metre}$}
    \label{fig:antenna}
\end{figure}

The bow-tie antenna (Figure \ref{fig:antenna}) utilized in this study is based on design constrained in size, with dimension of less than \SI{30}{\milli\metre} by \SI{13}{\milli\metre}, thus ensuring comfortable wearability.
Despite the antenna structure being of a reduced size, return loss of less than $<\SI{-5}{dB}$  has been achieved (see Fig.~\ref{fig:s11}).
The excitation of the field is predominantly linear polarised, and is focused mainly in the direction of the vocal tract as a result of the high dielectric constant of the tissue.
The antenna structure is fabricated from a \SI{0.8}{\milli\metre} thick FR4 printed circuit material with \SI{35}{\micro\metre} copper cladding. For the purpose of simulation, the structure is directly excited by a lumped power source. The antenna used in the measurement has an additional balance to unbalanced (balun) transformer, to ensure symmetric feeding of the two bow-tie arms out of an asymmetric coaxial cable. 

\subsection{Scattering Matrix Measurement}

\begin{figure}
    \centering
    \includegraphics[scale=1.1]{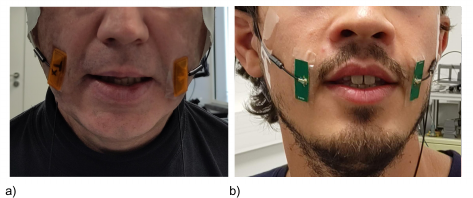}
    \caption{a) Subject~1 and b) Subject~2 pronouncing the vowel /i/ with the antenna attached. In a), the antennas appear orange because they are covered with a protective cap. This precaution was later deemed unnecessary, so the cap is not present in b). Since the caps had no measurable effect on antenna performance, the electrical parameters are identical in both configurations.}
    \label{fig:exp_setup}
\end{figure}

For Subject~1, the measurements were conducted at TU Dresden using a VNA (Rohde \& Schwarz ZVL6) .
For Subject~2 the measurements were performed subsequently at the university of Rennes using a VNA Keysight M9374A
mounted in a Keysight M9005A PXIe Chassi.
Prior to commencing the measurements, a full 2-port short open load thru (SOLT) calibration was performed. It is important to note that this calibration did not include micro-coaxial cables that are directly attached to the antenna, as their interface is not readily accessible. In order to compensate for the expected impedance transformation, the cable was deembedded based on a physical model of the micro coaxial cable. 

As illustrated in Fig.~\ref{fig:exp_setup}, the same subjects who underwent the MRI acquisition had the antennas attached to their cheeks and connected to the VNA which was operated at an output power of -10~dBm. 
The vicinity of the corners of the cheeks were identified as good locations for a phoneme recognition task \cite{Wagner2022, Possamai2022, Menezes2025} , and thus were selected for this study (see Fig.~\ref{fig:meshes}c).
The antennas were attached to the skin using biocompatible double-sided tape (3M 9889).
The seated participants were instructed to hold the desired phonemes, /a, i, u/, in a stable manner while the scattering matrix was measured. Pictures were taken from the front and both sides while they pronounced the phonemes.

To account for inherent variability due to articulation differences and other experimental factors, the scattering matrix measurements were repeated three times for each phoneme. The measurements were performed in the frequency range of 1-6 GHz, similar to the methodology used by Wagner et al. \cite{Wagner2022}. This frequency band was selected to encompass all practically usable frequencies, as radar-based sensors benefit from the widest possible bandwidth. The lower frequency limit is constrained by the physical size of the antenna, while the upper limit is determined by signal attenuation.

\subsection{Surface Mesh Generation}\label{sec:surf_mesh_generation}

To simulate wave propagation in the head, a geometrical description of the various anatomical parts is required. This is achieved by using closed surfaces that correspond to the boundaries of specific tissues. These surfaces are represented using surface meshes, which are collections of interconnected triangles. The goal of this particular stage in the process was to generate these meshes from the segmented MRI data.

Surface meshes generated using standard free tools, such as Meshlab or FreeCAD, were not accepted by the version of HFSS used (2022 R2). Mesh errors were detected when importing these meshes into HFSS, which were not found in other softwares. However, a test run using the C++ computational geometry library CGAL (version 5.4.1) \cite{cgal:eb-24b} demonstrated that the mesh obtained was considered error-free by HFSS. A custom C++ program, freely available on GitHub \cite{surf_mesh_from_MRI},
was developed for this purpose.

Segmented MRI data were converted to surface meshes using the CGAL, Eigen3 (version 3.4.0) \cite{eigen_manual}, and Visualization Toolkit (VTK) (version 9) \cite{VTK_textbook_2006} C++ libraries. 
A minimalist user interface was implemented with the Qt5 C++ graphical interface library (version 5.15.3), which takes segmentation masks in Neuroimaging Informatics Technology Initiative (NIfTI) format as input. 
Users can specify meshing parameters, including the maximal edge dimension in millimeters, the number of smoothing iterations, and the percentage of simplification. 
The surface mesh is generated from the segmentation mask using the \emph{make\_surface\_mesh} function from the "3D Surface Mesh Generation" package of CGAL \cite{Rineau2007}. Typically, an edge dimension of 1~mm is requested, but the resulting surface mesh is not smooth as it follows the voxels of the mask. To address this, the mesh is smoothed using the \emph{smooth\_shape} function from the "Polygon Mesh Processing" CGAL package, which is based on the curvature flow algorithm \cite{desbrun1999, kazhdan2012}. Usually, six iterations are performed, as this was found to be a good trade-off between removing voxelization artifacts and avoiding over-smoothing. However, the resulting surface mesh has an unnecessarily high number of elements, which would slow down FEM simulations without improving accuracy. To mitigate this, the number of elements is reduced using the \emph{edge\_collapse} function from the "Triangulated Surface Mesh Simplification" CGAL package, typically retaining 10\% of the vertices. For example, this results in about 30,000 faces for the air/tissues interface.

It is worth noting that a similar final number of vertices could have been achieved by initially requesting a larger edge dimension. However, this approach would have resulted in the loss of finer details, such as teeth, and the voxelization artifacts would not have been adequately compensated. The proposed method, on the other hand, allowed for the generation of meshes with increased density in areas requiring a more detailed description, while simultaneously reducing the number of triangles in smoother surfaces, such as cheeks.

However, this procedure can lead to the creation of flat parts, where two faces are very close together, defining a volume with a very small thickness. This issue is particularly prevalent in intricate surfaces, such as teeth or brain structures, and is not suitable for simulations, as it can generate errors in HFSS. To mitigate this problem, a higher number of vertices was sometimes used, and/or a larger edge dimension was requested.

\subsection{FEM Model Creation}

\begin{figure}
    \centering
    \includegraphics[scale=1.1]{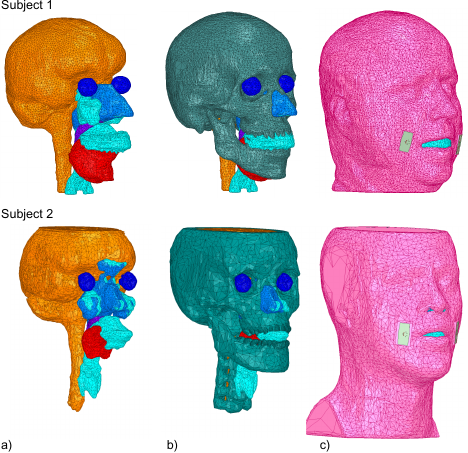}
    \caption{Surface meshes generated in HFSS for the vowel /a/, top row shows Subject~1, and bottom row Subject~2. a) From top to down: brain tissues, eyes, nasal mucosa, internal air and tongue, b) same with skul and teeth, c) overall envelope with the antennas.}
    \label{fig:meshes}
\end{figure}

Surface meshes were generated using the method previously described in Section~\ref{sec:surf_mesh_generation}. For this purpose, some masks corresponding to the same material, such as bones, were merged in order to have one surface mesh per simulated material. The resulting surface meshes are illustrated in Fig.~\ref{fig:meshes}.
They were then imported via the modeler of HFSS. Initially, the volume enclosed by the outer surface, primarily the skin, was defined. Subsequently, surface meshes for various internal elements, including the brain, bones, eyes, nasal mucosa, tongue, soft palate, and air, were created.
To avoid overlap, these surface meshes were first subtracted from the other intersecting elements using boolean operations. This process created voids that matched exactly the shape of each element, ensuring no gap at the boundary between the various materials when the internal parts were added.

The relevant tissue properties were assigned to the volumes enclosed by the different surface meshes. These properties were specified in terms of frequency-dependent relative permittivity and conductivity. They were obtained from a parametric model based on measurements provided by the web application \cite{web_app_prop_tissues}
of the Istituto di Fisica Applicata "Nello Carrara" (IFAC) \cite{gabriel1996_1, gabriel1996_2, gabriel1996_3}. The tissues considered included white matter, gray matter, cerebrospinal fluid, cortical bone, nasal mucosa, eye vitreous humor, and muscles. Initially, fat was considered for other non-segmented tissues in the envelope. However, initial results indicated that this did not match the order of magnitude of the measurements. Considering these tissues as muscle in a first approximation proved to be more realistic. 
Nevertheless, better estimations of the material properties could be made in the future (see discussion section~\ref{sec:discussion}).

Once the different anatomical parts were assembled, the antennas were positioned. The antenna model was directly imported from its computer aided design (CAD) model via the HFSS modeler, although it was simplified compared to the actual devices. The balun was removed, retaining only the copper sheets and the printed circuit board (PCB), as the influence of the balun and connecting cables was considered minimal and unnecessarily increased the complexity of the modeling task.

Given that the antennas were attached with adhesive to the cheeks and the distance between the antenna and the cheek was expected to play an important role, it was crucial to place the antennas as close as possible to the cheek and ensure this distance was uniform. To fulfill both conditions, parallelepipeds with slightly larger dimensions than the antennas were created and placed at the surface of the cheeks. These parallelepipeds were fused with the head model using a boolean operation. This created locally perfectly flat surfaces in front of which the antenna models could be brought as close as desired without intersecting the head model, thus avoiding galvanic contact.

The location of the antennas was adjusted to correspond as closely as possible to the pictures taken during the experiment. This task was challenging because the 3D head geometry may have varied slightly between the experiment and the MRI acquisition, due to changes in position (supine vs. seated) and the natural variability of phoneme production. Finding a correspondence between the pictures and the 3D models was also difficult.

A parametric analysis showed that a distance of \SI{1.}{\micro\metre} from the head model ensured a good match with the experimental data, specifically achieving an $S_{11}$ of the same order of magnitude (see Fig.~\ref{fig:s11}).

Lumped ports were defined between the two sides of the bow-tie antennas. Simulations were conducted from 1~GHz to 6~GHz using an interpolating sweep. 
The SAR was computed following the ICNIRP guidelines (standard IEC/IEEE 62704-4) using the CertificationSAR algorithm provided in the field calculator in HFSS at 1~GHz, 3~GHz and 6~GHz with an input power of 0.5~W.

\section{Results}

\begin{figure}
    \centering
    \includegraphics[scale=1.1]{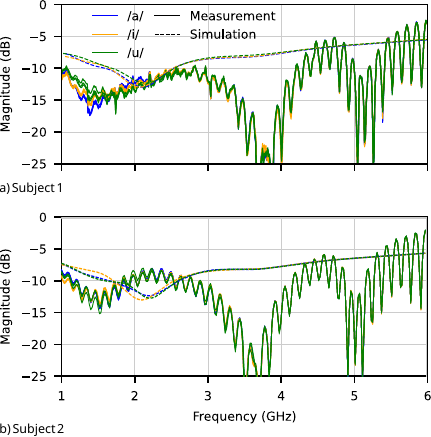}
    \caption{Reflection coefficient measured and simulated for three different vowels. 
	The oscillations in the measured data are due to imperfect deembedding.
	}
    \label{fig:s11}
\end{figure}

\begin{figure}
    \centering
    \includegraphics[scale=1.1]{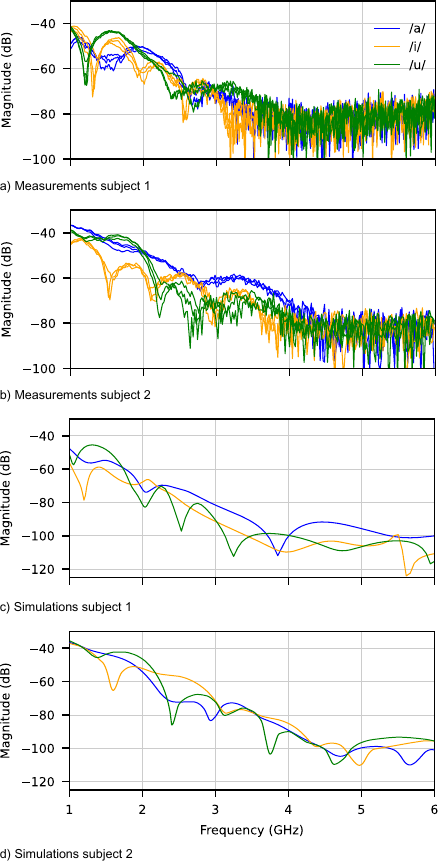}
    \caption{Transmission coefficient measured and simulated for three different vowels.}
    \label{fig:s12_comp}
\end{figure}

\begin{figure}
    \centering
    \includegraphics[scale=1.1]{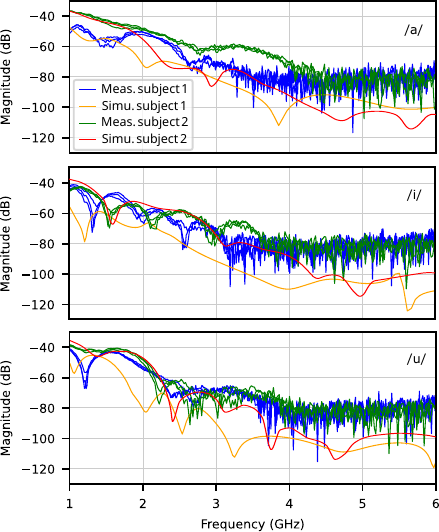}
    \caption{Transmission coefficient measured and simulated for three different vowels presented separately for each vowel.}
    \label{fig:s12}
\end{figure}

\begin{figure*}
    \centering
    \includegraphics{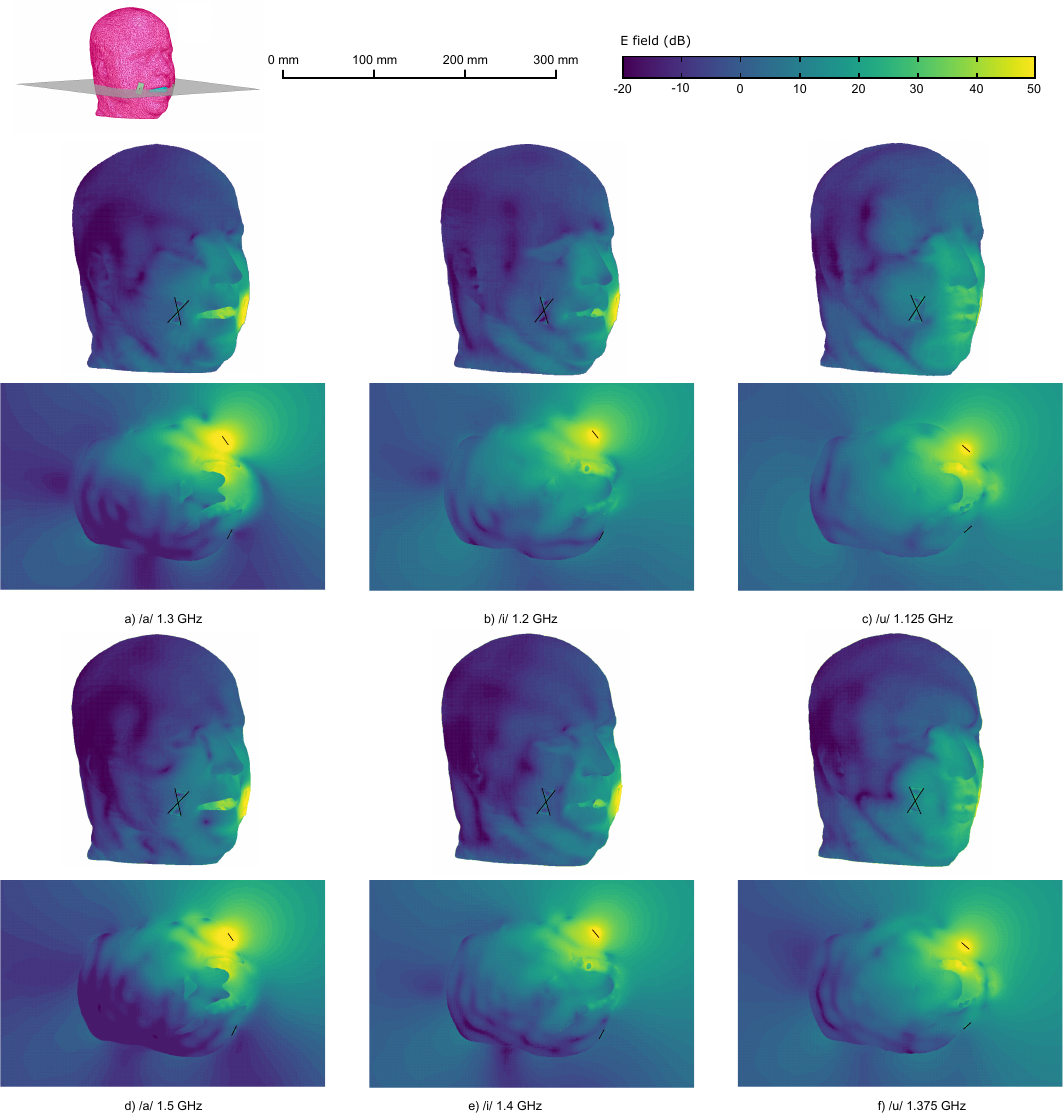}
    \caption{Magnitude of the electric field on the surface of the head of Subject~1 and in a horizontal cut (shown on top of the figure) for the vowels /a, i, u/ at frequencies corresponding roughly to a), b), c),  the first dip of the transmission coefficient shown in Fig.~\ref{fig:s12} and d), e) f), the following maximum.
    }
    \label{fig:E-field1}
\end{figure*}

\begin{figure*}
    \centering
    \includegraphics{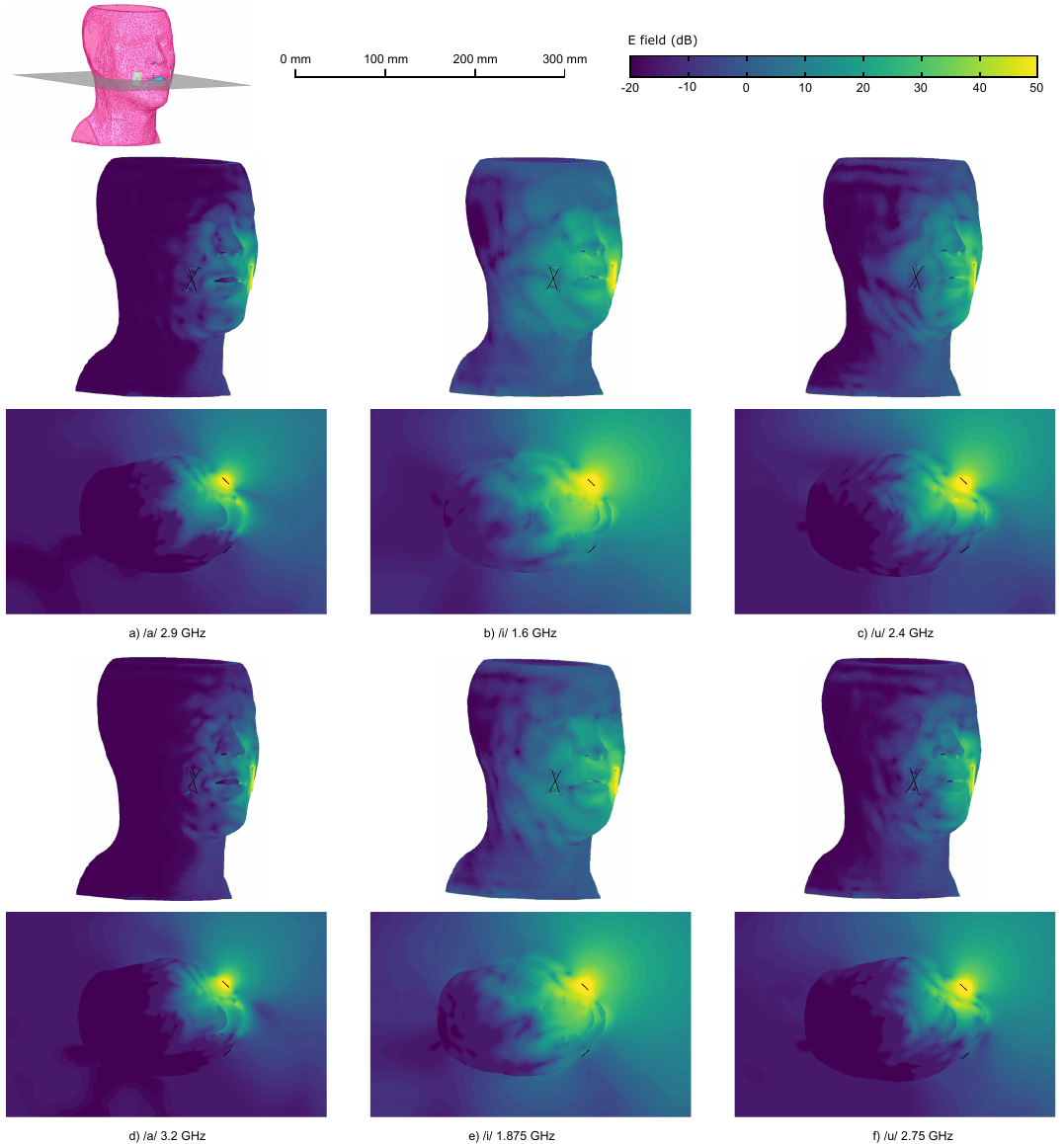}
    \caption{Magnitude of the electric field on the surface of the head of Subject~2 and in a horizontal cut (shown on top of the figure) for the vowels /a, i, u/ at frequencies corresponding roughly to a) and b), the first dip and c) the second dip of the transmission coefficient shown in Fig.~\ref{fig:s12} and d), e) f), the following maximum.
    }
    \label{fig:E-field2}
\end{figure*}

\begin{figure*}
    \centering
    \includegraphics{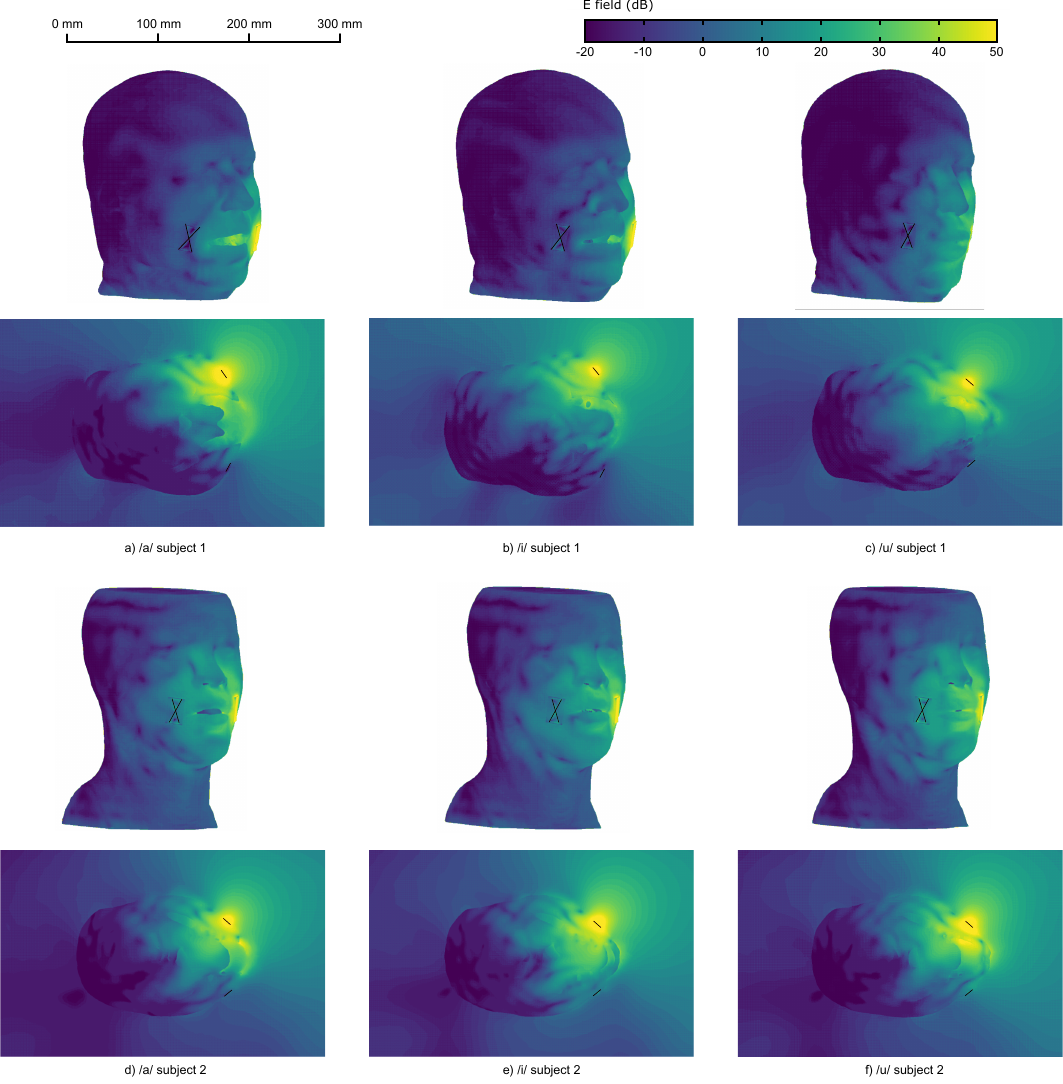}
    \caption{Magnitude of the electric field on the surface of the head and in a horizontal cut (shown on top of Figs.~\ref{fig:E-field1} and \ref{fig:E-field2}) for the vowels /a, i, u/ at 2~GHz.
	}
    \label{fig:E-field_same_freq}
\end{figure*}

The reflection coefficient, as shown in Fig.~\ref{fig:s11}, exhibits oscillations in the measured data due to imperfect deembedding. Specifically, the effect of standing waves in the cable connecting the network analyzer to the antenna was not fully compensated. Despite this, the measured and simulated reflection coefficients are of the same order of magnitude. However, the simulated reflection coefficient is slightly higher, around --7.5 dB at 1 GHz, instead of --10~dB for the measurement.
Both simulated and measured reflection coefficients exhibit one dip around 2~GHz and 1.5~GHz respectively. However, the measured reflection coefficient has two prominent dips at higher frequencies which are not seen in the simulated one.
The differences between the reflection coefficients for various vowels are minimal, reaching up to 2 dB near the minimum (around 1.5 GHz for experimental data and 2 GHz for simulations). Notably, these differences are marginally larger in the experimental data for subject 1 compared to the simulations.

Fig.~\ref{fig:s12_comp} shows the measured and simulated transmission coefficients for all vowels. Fig.~\ref{fig:s12} separately displays the simulated coefficients alongside the measured ones for each vowel, as combining them in one plot would be unclear.

Overall, the measured transmission coefficients decrease with increasing frequency and become very noisy above 3~GHz. 
Distinct minima are observed at specific frequencies for each vowel and subject:
\begin{itemize}
  \item \textbf{Vowel /a/}
    \begin{itemize}
      \item Subject 1: 1.5~GHz, 2.7~GHz
      \item Subject 2: 2.8~GHz
    \end{itemize}
  \item \textbf{Vowel /i/}
    \begin{itemize}
      \item Subject 1: 1.3~GHz, 1.9~GHz, 2.6~GHz, 3.2~GHz
      \item Subject 2: 1.5~GHz, 2.1~GHz, 2.9~GHz
    \end{itemize}
  \item \textbf{Vowel /u/}
    \begin{itemize}
      \item Subject 1: 1.2~GHz, 2.4~GHz
      \item Subject 2: 1.3~GHz, 2.2~GHz, 2.7~GHz
    \end{itemize}
\end{itemize}

The transmission coefficient obtained from simulations is generally of the same order of magnitude as the measured data up to approximately 2–3~GHz, except for the vowel /i/ of Subject~1, for which it is about 10 dB lower across the entire studied frequency range. 
Among the three vowels, /u/ exhibits the highest transmission magnitude in the 1–2 GHz range, with values around -45~dB at approximately 1.5~GHz for Subject~1 and -35~dB at 1~GHz for Subject~2. 
Both the measured and simulated transmission coefficients show similar overall trends, characterized by a global decay with increasing frequency and the presence of local minima. 
However, for Subject~1 and for vowel /a/ of Subject~2, the simulated curves are shifted toward lower frequencies compared to the measurements, whereas for vowels /i/ and /u/ of Subject~2, the simulated results are slightly shifted toward higher frequencies. This behavior is more clearly illustrated by the positions of the minima for each vowel and subject:
\begin{itemize}
  \item \textbf{Vowel /a/}
    \begin{itemize}
      \item Subject~1: 1.3~GHz, 2.0~GHz, 3.9~GHz
      \item Subject~2: 2.95~GHz, 4.7~GHz, 5.7~GHz
    \end{itemize}

  \item \textbf{Vowel /i/}
    \begin{itemize}
      \item Subject~1: 1.2~GHz, 1.9~GHz, 4.0~GHz, 5.1~GHz, 5.6~GHz
      \item Subject~2: 1.6~GHz, 3.2~GHz, 4.4~GHz, 4.95~GHz
    \end{itemize}

  \item \textbf{Vowel /u/}
    \begin{itemize}
      \item Subject~1: 1.05~GHz, 2.0~GHz, 2.5~GHz, 3.2~GHz, 4.7~GHz, 5.9~GHz
      \item Subject~2: 1.4~GHz, 2.4~GHz, 3.1~GHz, 3.8~GHz and 4.7~GHz
    \end{itemize}
\end{itemize}

There is limited variability among the three measurements for each vowel, typically around 5~dB, except for certain specific repetitions or frequencies where the variability can reach up to 10~dB. Notably, one of the measurements of Subject~1 for the vowel /u/ shows the second minimum at a higher frequency, approximately 2.7~GHz instead of 2.4~GHz. 
Increased variability is also observed for vowel /u/ of Subject~2 above 2~GHz.
The variability is more pronounced near the minima, such as the minimum around 1.5~GHz for the vowel /a/ of Subject~1.

There are significant differences between the vowels, with variations up to 15~dB between /a/ and /u/ around 1.5~GHz for Subject~1, and between /i/ and /u/ around 1.5~GHz for Subject~2. 
The minima for each vowel occur at different frequencies and exhibit varying degrees of sharpness. 
Moreover, the number of minima differs, with /i/ showing the largest number.
Similar differences are observed in the simulations. 
However, above 3~GHz, the minima cannot be compared with the measurements because of excessive noise in the experimental data. In addition, for Subject~1, the third and fourth measured minima of the vowel /i/ do not correspond to  simulated ones. Similarly, for Subject~2, the second measured minimum of the vowel /i/ has no corresponding minimum in the simulated data.

Most of the differences between the transmission coefficients of the two subjects are observed for vowel /a/.
The transmission coefficient of Subject~2 is substantially higher, exceeding that of Subject~1 by more than 10~dB, except in the 1.5–2.5~GHz range for the experimental data and in the 2–3~GHz range for the simulation. The first two minima observed for Subject~1 are not present for Subject~2. More generally, across all vowels, certain minima appear for one subject but are absent for the other. When corresponding minima can be identified, they occur at different frequencies: for vowel /i/, the minima of Subject~2 are shifted toward higher frequencies relative to those of Subject~1, whereas for vowel /u/, the minimum above 2~GHz occurs at a lower frequency for Subject~2. Despite these differences, there are also remarkable similarities, particularly in the overall shape and amplitude for vowels /i/ and /u/. The vowel /i/ exhibits regularly spaced minima and maxima, while /u/ shows a pronounced amplitude drop above 2~GHz and minima occurring at similar frequencies for both subjects, even though the first minimum of Subject~2 is barely visible (more clearly observed in the simulations). Even for vowel /a/, a minimum near 3 GHz is observed for both subjects.

The magnitude of the simulated electric field is presented in Figs.~\ref{fig:E-field1}, \ref{fig:E-field2} and \ref{fig:E-field_same_freq}, both on the surface of the head and on a horizontal cross-section containing the antennas. 
The frequencies selected in Figs.~\ref{fig:E-field1} and \ref{fig:E-field2} correspond to minima and maxima of the transmission coefficient for each vowel and subject, as shown in Fig.~\ref{fig:s12}

The electric field is more intense in bones and internal air volumes, which can be attributed to the differing impedances and losses of these media. 
On the surface of the head, sharp local minima and maxima are observed at the antenna locations due to the metal-skin interface.
Additionally, local minima and maxima can be seen on the skin's surface, both inside and outside the head. These patterns can be interpreted as Mie scattering~\cite{NikolayevPRL}, which is relevant given the proximity of the wavelength (ranging from about 30 cm to 5 cm in the air) to the dimensions of the head (approximately 20 cm). 
For instance, in the case of Subject~1, three local minima may correspond to a quadrupole pattern that appears distorted due to the asymmetry and complex geometry of the head (see Figs.~\ref{fig:E-field1}a, \ref{fig:E-field1}d and \ref{fig:E-field_same_freq}b).
For the other fields, local minima are also observed in the radiated field, though they do not exhibit any clear relationship to specific Mie scattering modes.
One can also see that the minima are more widely spaced outside the head than inside. This aligns with the shorter wavelength in head due to higher permittivity of the head tissues.
Thus, Mie scattering modes may underlie simplified modeling approaches to this phenomenon.

Each minimum in the transmission coefficient corresponds to a local minimum on the skin surface at the location of the receiving antenna. This is shown by the surface field in Figs.~\ref{fig:E-field1}a-c, Figs.~\ref{fig:E-field2}a-c and Figs.~\ref{fig:E-field_same_freq}a and \ref{fig:E-field_same_freq}c. Conversely, the maxima in the transmission coefficient match the local maxima of the electric field at the receiving antenna's location, as seen in Figs.~\ref{fig:E-field1}d-f, Figs.~\ref{fig:E-field2}d-f, Fig.\ref{fig:E-field_same_freq}b and Figs.\ref{fig:E-field_same_freq}e-f.

The local minima on the surface can sometimes correspond to internal dips, as seen for /a/ in Fig.~\ref{fig:E-field2}a, /i/ in Figs.~\ref{fig:E-field1}b and \ref{fig:E-field2}b and /u/ in Figs.~\ref{fig:E-field2}c and \ref{fig:E-field_same_freq}c, or to external minima, as seen for /a/ in Fig.~\ref{fig:E-field1}a. In some cases, such as for /u/ in Fig.~\ref{fig:E-field1}c and /a/ in Fig.~\ref{fig:E-field_same_freq}a, these dips result from a combination of low amplitudes both inside and outside. Notably, the local minima for /i/ of Subject~1 and /u/ of Subject~2 are particularly sharp, reflecting the sharpness of their transmission coefficient minima.

The local maximum of the transmission coefficient of vowels /i/ and /u/ of Subject~1 is associated with an internal maximum, as depicted in Figs.~\ref{fig:E-field1}e-f.
For the vowels /a/ of Subject~1 and /i/ of Subject~2, the local maximum is associated with an internal maximum that coexists with a low external amplitude, as depicted in Figs.~\ref{fig:E-field1}d and \ref{fig:E-field2}e. 
A similar observation is noted at 2 GHz with /i/ of Subject~1 in Fig.~\ref{fig:E-field_same_freq}b, although this does not precisely align with the 2.1 GHz local maximum in the transmission coefficient.
A slightly reversed trend is observed for vowel /a/ of Subject~2, where the internal field is marginally weaker than the external field (see Fig.~\ref{fig:E-field2}d).
However, for /u/ of subject 2, the electric field exhibits high amplitudes both internally and externally, as shown in Fig.~\ref{fig:E-field2}f.
Swapping the roles of the receiving and transmitting antennas yields similar observations.

At a given frequency, 2~GHz in Fig.~\ref{fig:E-field_same_freq}, the pattern of local minima and maxima changes depending on the vowel being articulated. In certain configurations and frequencies, these patterns can become sharper, as observed for the vowel /u/ in Fig.~\ref{fig:E-field_same_freq}c and \ref{fig:E-field_same_freq}f.
The electric field distributions in the regions behind, above, and to the right of the receiving antenna tend to be more similar across different vowels for each specific subject.
There are substantial differences between the two subjects. For instance, Subject~2 does not exhibit sharp local minima near the tip of the nose. Nonetheless, a few similarities can be identified--for example, a local minimum appears at the corner of the lips for vowel /i/, as shown in Figs.~\ref{fig:E-field_same_freq}b and \ref{fig:E-field_same_freq}e.
In the internal field, similar low amplitude lines aligned with the cheeks can also be identified, in particular for /i/ in Figs.~\ref{fig:E-field1}b and \ref{fig:E-field2}b.

The SAR averaged over 10~g of tissue increases with frequency, rising from approximately 17~W/kg at 1~GHz to a maximum of about 40~W/kg at 6~GHz, with only minor variations between vowels and subjects. Similarly, the SAR averaged over 1~g of tissue increases from around 51~W/kg at 1~GHz to approximately 280~W/kg at 6~GHz. These relatively high values, compared to those observed in typical applications such as mobile phones, can be attributed to the direct contact between the antenna and the skin (albeit with an adhesive layer in between).
To comply with the ICNIRP limit of 2~W/kg, the input power should remain below 0.025~W (14~dBm), while to satisfy the more stringent U.S. limit of 1.6~W/kg, it should remain below 0.0025~W (4~dBm). The -10~dBm power used in the experiments is therefore well within both safety limits. This allowable power could be further increased by introducing a dielectric matching layer with an intermediate permittivity between air and soft tissue, as done in \cite{Soares2022}.

\section{Discussion}\label{sec:discussion}

Differences between experimental results and simulations are expected for several reasons. Firstly, the cables and balun are not included in the simulations, which is anticipated to primarily affect the reflection coefficient. Despite this omission, the agreement for the reflection coefficient can be considered good, as the general order of magnitude is similar. Simulating the cables would introduce ripples akin to the artifacts observed in the experimental data. Thus, the antenna-skin interface appears to be accurately simulated.

Other sources of deviation between experiments and simulations include 
differences in the imaged volume (e.g. the neck was excluded for Subject~1, whereas the top and sides of the head were cropped for Subject~2; see Fig.~\ref{fig:meshes}),
segmentation inaccuracies and/or incompleteness (e.g., neck bones were not segmented for Subject~1), approximations in material properties, discrepancies in antenna placement between simulations and experiments, differences in subject positioning between MRI acquisition and scattering matrix measurement (supine vs. seated)\cite{Stone2007}, the noise of the MRI machine potentially affecting vowel pronunciation (Lombard effect \cite{Zollinger2011}), and natural variations in articulation.

Despite these numerous sources of deviations, the simulated transmission coefficient closely matches the measured data up to 3~GHz, exhibiting identical minima and maxima at similar frequencies and magnitudes. 
Thus, there is surprisingly good qualitative agreement.

However, above 3 GHz, a comparison is not feasible due to the low signal-to-noise ratio of the experimental data.
This results from stronger attenuation, caused by increasing dielectric and conductive losses at higher frequencies.
Possible mitigation strategies include increasing the transmitted power (within SAR limits), 
enhancing receiver sensitivity through longer measurement times, reducing the intermediate frequency bandwidth of the VNA, or using antennas with higher sensitivity.
Alternatively, shorter transmission paths could be achieved by placing antennas inside the nose, mouth, or ear cavities. Wavefront shaping could also be employed to optimize the transmission path
\cite{Soares2025}.

The overall intersubject difference in the simulated transmission coefficient amplitude can be attributed to variations in head size and geometry. Specifically, the average distance between the antennas is 79 mm for Subject 2, compared to 104 mm for Subject 1. Furthermore, the average relative angle between the antennas is 78° for Subject 2, while it is 105° for Subject 1. As a result, the antennas in Subject 1 are positioned farther apart and are less directly facing each other, which explains the lower transmission coefficient observed in the simulations.
In contrast, the experimental data show that the transmission coefficient amplitudes for the vowels /i/ and /u/ are similar for both subjects. This suggests that differences in tissue properties—identical in the simulations—compensate for the geometric discrepancies in the real measurements.

For Subject~1, the shift of the simulated shape toward lower frequencies is likely caused by an overestimation of the average relative permittivity of soft tissues. In contrast, the opposite trend is observed for Subject~2 /i/ and /u/, probably resulting from an underestimation of this parameter. Such intersubject differences may arise from general anatomical variations, differing fat-to-muscle ratios, or incomplete segmentation—particularly the omission of fat and other tissues such as blood vessels. Since variations in the average relative permittivity of soft tissues between subjects can induce the observed frequency shifts, including this parameter in the analysis could make speech recognition from transmission spectra more reliable. Specifically, estimating and compensating for this parameter may help correct frequency shifts, similar to existing approaches in speech processing that compensate for intersubject variations in vocal tract length. 
Another implication of these frequency shifts potentially related to intersubject tissue properties variation is that broad band signals are necessary to capture similar patterns.
However, this trend requires further confirmation, as it was not observed for the vowel /a/ of Subject 2. It should be noted that the agreement between experiment and simulation was also poorer for this particular phoneme and subject.

The reflection coefficient provides limited information regarding articulatory movements, as the differences between vowels are too subtle to easily detect them. Without specific signal processing, the reflection coefficient is of little use for this purpose. 
This is further emphasized by the similarity observed between both subjects: the anatomical differences have little effect, indicating that it is primarily determined by interactions with tissues located very close to the antenna.

In contrast, the relatively small variation between the transmission coefficients of different repetitions, coupled with larger differences between vowels, indicates that they can be distinguished in a reproducible and reliable manner.
Thus, the transmission spectrum likely contains more information about articulatory movements than the reflection spectrum. This aligns with observations from \cite{Birkholz2018}, who noted that the transmission spectrum provided the most information for phoneme classification. 

The slightly larger variation observed in one repetition of /u/ of Subject~1 is probably due to a larger variation in articulation compared to the other repetitions.
Recording the sound could help evaluate the relationship between articulatory variation and vowel quality by comparing variations in the transmission coefficient with formants
variations. 
In speech science, formants are peaks in the spectrum of the acoustic signal, typically corresponding to acoustic resonances in the vocal tract. They play a crucial role in phoneme identification.

The fact that variability mostly occurs in the minima may be because these regions are more sensitive to small changes in articulatory position. Therefore, they may provide accurate information regarding small articulatory variations, while more stable features may offer robust information about the global configuration, aiding in phoneme identification.

The global amplitude decay with frequency can be understood as a complex interplay of several factors, including the increase in tissue conductivity with frequency, the antenna's radiation efficiency, and various reflections at material interfaces \cite{Nikolayev2018}. 

The gaps and maxima in the transmission coefficient result from variations in resonance patterns with frequency and articulatory configuration. When the antenna location coincides with a minimum of a resonance pattern, it induces a gap, and vice versa. This interaction occurs at both the receiving and sending antenna locations.

A specific minimum or maximum in the transmission coefficient can be, in some cases, solely related to a resonance pattern localized within the head (Figs.~\ref{fig:E-field1}b, \ref{fig:E-field1}e-f,  and \ref{fig:E-field2}a-c), in other cases, solely to a resonance pattern occurring outside the head (Figs.~\ref{fig:E-field1}a and \ref{fig:E-field2}d), or to a combination of both types of resonances (Figs.~\ref{fig:E-field1}c, \ref{fig:E-field1}d and \ref{fig:E-field2}e-f). In the latter case, the combination of both can induce the same effect at the same frequency (e.g., a minimum in the case of /u/ of Subject~1 in Fig.~\ref{fig:E-field1}c, or a maximum in the case of /u/ of Subject~2 in Fig.~\ref{fig:E-field2}f) or have antagonistic effects, with one being slightly stronger than the other at certain frequencies.
The combination of antagonistic effects in the case of /a/ may explain its smoother transmission coefficient, where, for Subject~1, the first minimum and maximum are less pronounced and spread over a wider frequency range, and for Subject~2, a smooth transmission coefficient is seen until about 2.5~GHz. 

From this perspective, it is clear that modifying the antenna location can substantially alter the transmission coefficient, inducing minima and maxima at different frequencies. Therefore, antenna location is a key parameter for SSI, as highlighted by \cite{Possamai2022, Wagner2022, Menezes2025}.

It is also evident that anatomical differences strongly influence the field distribution patterns, as illustrated in Fig.~\ref{fig:E-field_same_freq}. As expected, as a consequence of this, substantial intersubject variations can be observed in the transmission spectra, shown in Figs.~\ref{fig:s12_comp} and \ref{fig:s12}.  However, it is remarkable that, despite the numerous sources of variation between the two subjects, the transmission spectra of the different vowels share common characteristics—particularly for /i/ and /u/. Although similarities in the field distributions are more difficult to discern, this suggests that the articulatory configuration affects electromagnetic wave transmission in a comparable manner. In fact, parallels can be drawn between how the acoustic signal and the transmission spectrum encode phonemic information: despite multiple sources of variation (anatomy, articulatory differences, etc.), similar patterns are produced. These patterns, though somewhat distorted relative to each other, still convey equivalent information, allowing a trained neural network or the human brain to recognize them. This observation explains why proof-of-concept studies of SSI based on variations in electromagnetic transmission spectra have proven successful and robust to inter-subject variability\cite{Birkholz2018, Wagner2022, Menezes2025}.
This also suggests that a speaker’s identity could be inferred from the transmission spectra, much like from the speech signal. Another implication is that broadband signals are necessary to develop applications robust to inter-subject variability.

Finding a relationship between the geometrical variations induced by articulation and the transmission coefficient is not straightforward because the resonance patterns are very sensitive to small geometrical changes. However, one can discuss general hypotheses.

One notable observation is that the mouth opening seems to affect the scattering modes. It may force some minima to be located near the receiving antenna, as observed for the vowel /a/, for which this effect seems more prominent. Additionally, one can hypothesize a stronger coupling between internal modes and scattering modes through the mouth. Some energy transmitted through the cheek may be re-emitted and scattered through the mouth. This may, in turn, affect the internal modes, possibly decreasing their quality factor by increasing losses.
Note that scattering modes are also affected by changes in face geometry, such as jaw movements.

Under these considerations, one can understand that the torso, which is not modeled here, may have an influence. In particular, the reflection on the shoulders may affect the shape of the scattering modes. It may be relevant in the future to include the torso or at least the shoulders, as is done in speech and hearing acoustics, to simulate the diffraction by the head and torso \cite{Meyer2025}.

In the future, articulatory models such as VocalTractLab  \cite{Birkholz2013} and Artisynth \cite{Lloyd2012} could play a key role in clarifying how variations in the transmission coefficient relate to articulatory movements. By enabling more controlled, gradual, and fine-grained articulatory changes, these models would allow for a more precise observation of the impact of phoneme transitions. They could also facilitate the development of simplified yet more explanatory models. Moreover, by simulating a wider range of vocal tract configurations, these tools could be used for articulatory inversion, allowing one to infer the corresponding vocal tract configuration from transmission coefficients. This approach could have practical applications in speech therapy—by providing articulatory feedback to patients—or in foreign language education, by helping learners acquire phonemes absent from their native language.

Articulatory motion also induces small changes in the antennas' location and orientation, particularly for Subject~1.
A larger distance between the antennas may result in a globally lower transmission coefficient. For example, for Subject~1, the vowel /i/ has the largest distance between the antennas (114~mm) compared to /a/ (106~mm) and /u/ (91~mm). This may contribute to the overall lower transmission coefficient for /i/. Conversely, this may explain in part why /u/ has the highest amplitude for Subject~1. 
For Subject~2 the variation of distance with the vowel is much smaller reaching a maximum of 4.5~mm instead of 23~mm for Subject~1. No prominent amplitude differences are observed for this subject.

The orientation of the antennas with respect to each other can have a similar effect due to their directivity. The directivity pattern of the bow-tie antenna consists of two lobes located on either side of the plane containing the antenna. Therefore, the greater the deviation from a parallel configuration, the lower the transmission coefficient. For Subject~1, the angle between them is smaller (closer to parallel, which would be 0°) for /u/: 83° compared to 114° for /i/ and 119° for /a/. This may also contribute to the higher amplitude of /u/.
For Subject~2, similarly to the distance, the variations of orientation are smaller than for Subject~1: maximum 8° instead of 36°. 
The intersubject differences in the variation of the antennas’ relative positions can be attributed to both anatomical factors and differences in articulation. It is possible that Subject~1 exhibited a more pronounced articulation.

\section{Conclusion}

The agreement between simulations and measurements is satisfying, indicating that the method is promising for future investigations. In alignment with previous work, it is confirmed that the transmission spectrum carries most of the information related to articulatory movements. The electromagnetic field can be interpreted as a Mie scattering pattern, consisting in internal resonances as well as scattering modes.

The observed minima and maxima of the transmission coefficient can be understood as the consequence of an interplay between the antenna locations and the geometry of the resonance patterns. To a lesser degree, the small relative displacements of the antennas are also expected to play a role in the variations of the transmission coefficient.

Although some effects, such as the influence of the mouth opening on scattering modes, have been discussed, relating specific articulatory gestures to changes in the transmission coefficient remains challenging. 
This may be addressed in the future using an articulatory model (e.g., VocalTractLab \cite{Birkholz2013} or Artisynth \cite{Lloyd2012}). 

Despite the sensitivity of resonances to geometrical details, similar patterns were observed for the vowels pronounced by both subjects. This finding is consistent with the success of previous proof-of-concept studies, which reported positive results across multiple participants. However, as only two subjects were included in the present study, extending this work to a larger and more diverse sample—particularly including female subjects—would provide a better understanding of how inter-individual anatomical variations influence phoneme identification. Such an expansion could also contribute to making radar-based speech technologies more robust and less dependent on individual anatomical differences.

\section*{Acknowledgment}

The authors are very grateful to Jens Frahm and Dirk Voit of the Max Planck Institute for Multidisciplinary Sciences, Yves Laprie of the LORIA of Nancy and Pierre-André Vuissoz of the IADI of Nancy for acquiring the MRI data, and sincerely appreciated João Vítor Possamai de Menezes assistance with proofreading.

\begin{IEEEbiography}[{\includegraphics[width=1in,height=1.25in,clip,keepaspectratio]{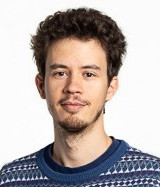}}]{R. Blandin}
was born in Pau, France, in 1989. He received his master’s degree in acoustics from the University Pierre and Marie Curie, Paris, France, in 2013, and his Ph.D. degree in acoustics from the University of Grenoble, France, in 2016. From 2019 to 2024, he worked as a Postdoctoral Researcher at the Dresden University of Technology (TU Dresden) on speech acoustics and silent speech interfaces based on radio frequencies. Since 2024, he has been working as a Postdoctoral Researcher at the University of Rennes on metasurface design for wavefront shaping. His research interests include waveguides, speech production, speech directivity, silent speech interfaces, bioelectromagnetics, and wave control in complex media.
\end{IEEEbiography}

\begin{IEEEbiography}[{\includegraphics[width=1in,height=1.25in,clip,keepaspectratio]{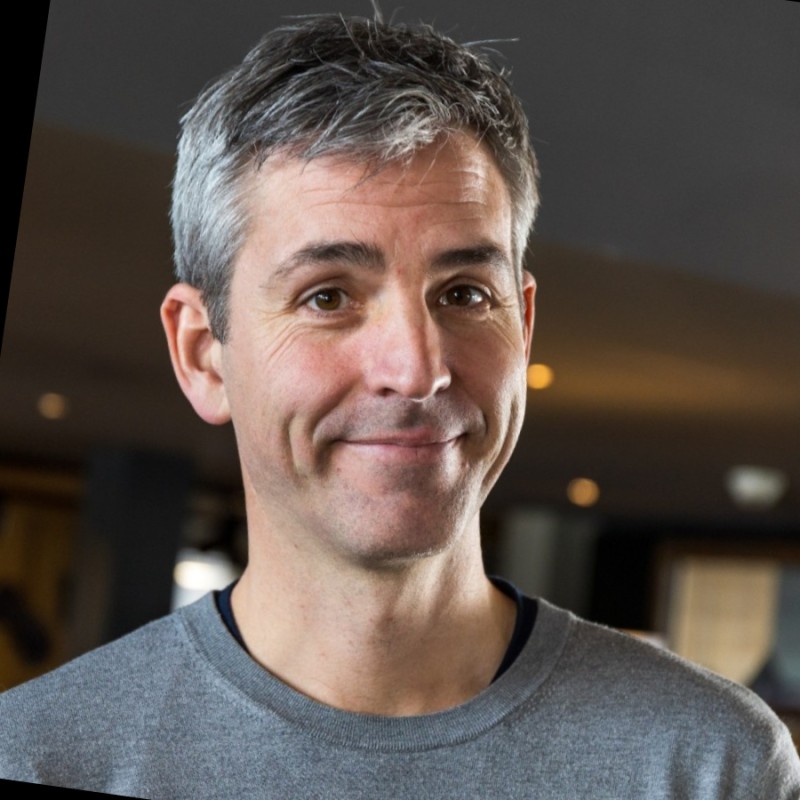}}]{B. Lloyd}
received an M.Sc. degree in electrical engineering from ETH Zurich, Zurich, Switzerland. He received a Ph.D. degree in electrical engineering from the Computer Vision Laboratory, ETH Zurich, Zurich, where his research focused on computational modeling of tumor growth and soft tissue simulation for virtual reality training simulators.
He is currently a Senior Scientist and Core Architect of the Sim4Life computational physics platform. His research interests include computational modeling, computational geometry, medical image analysis, deep learning, and image segmentation. He leads the Virtual Population team, developing state-of-the-art anatomical models of humans and animals for electromagnetic exposure simulations.
\end{IEEEbiography}

\begin{IEEEbiography}[{\includegraphics[width=1in,height=1.25in,clip,keepaspectratio]{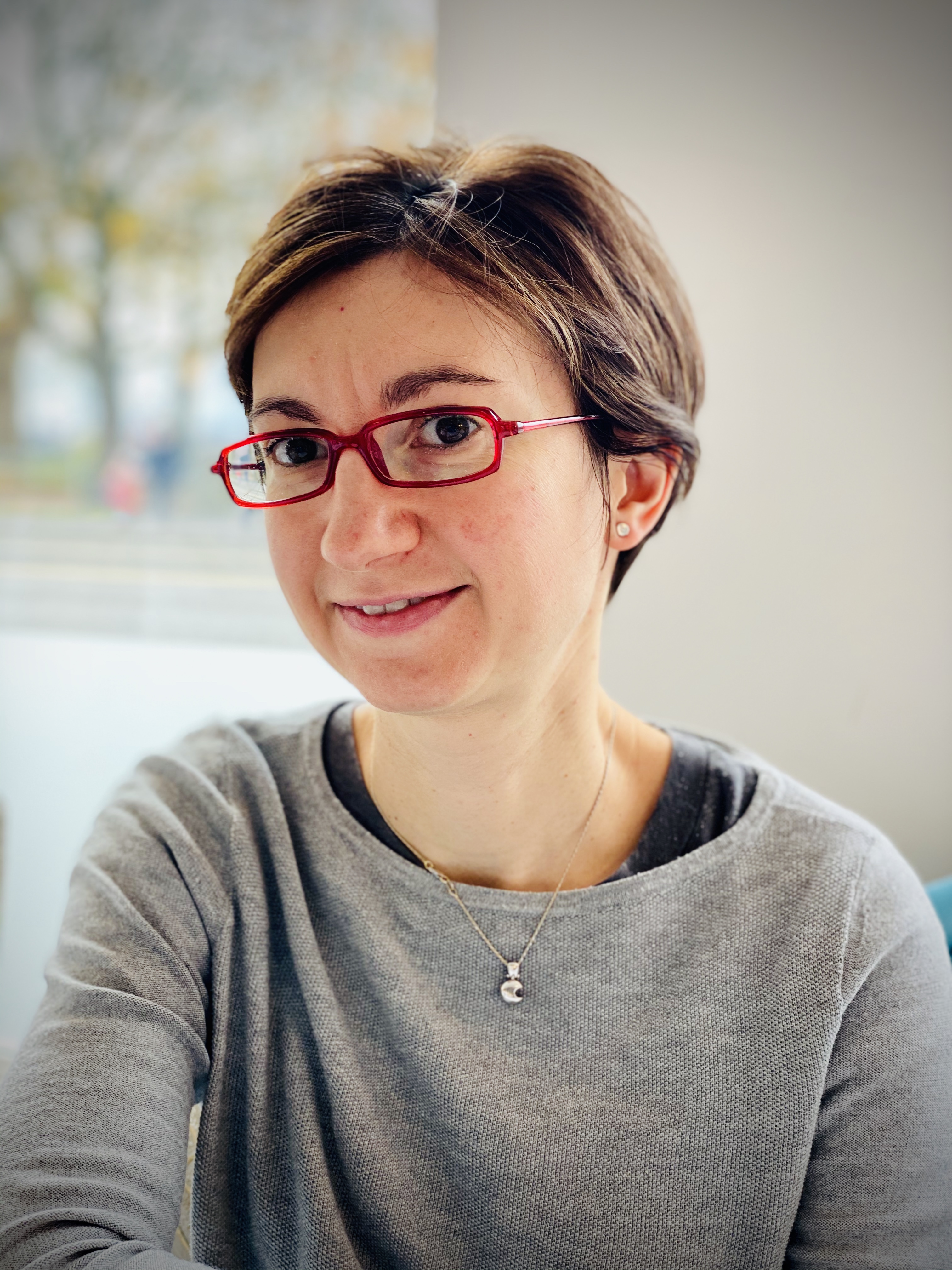}}]{S. Farcito}
received her degree in Biology in 2007 and her PhD in Neuroscience in 2012, both from the University of Torino, Italy. During her PhD, she spent two years at the University of Zurich investigating the role of neurotrophic factors during neuronal development.
Since 2012, she's been a scientific assistant at the IT’IS Foundation in Zurich, where she is involved in various stages of the generation of the Virtual Population models, IT’IS’ portfolio of high-resolution anatomical models for computational life science. Having worked with MR, CT and criosection images, she is an expert in image segmentation.
Her expertise also includes functionalizing the models with posing capabilities, placing active and passive implants within the models, as well as quality control of the anatomical models.
\end{IEEEbiography}

\begin{IEEEbiography}[{\includegraphics[width=1in,height=1.25in,clip,keepaspectratio]{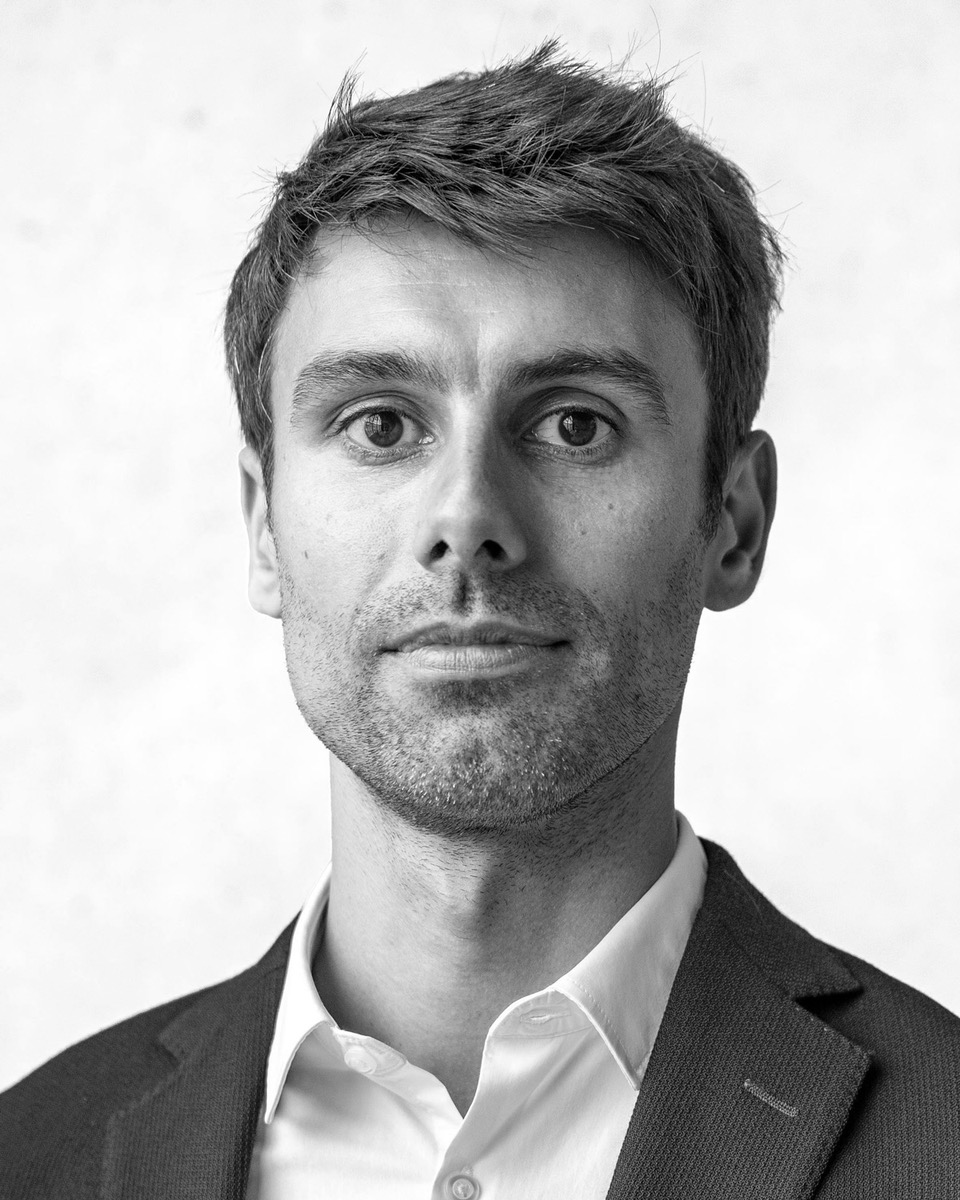}}]{D. Nikolayev (Senior Member, IEEE)}
received his M.Sc. degree summa cum
laude in applied electromagnetics from Lviv Polytechnic National University, Ukraine, in
2008. He earned dual Ph.D. degrees in electronics from the IETR laboratory of the
French National Center for Scientific Research (CNRS), France, and in electrical
engineering from the University of West Bohemia in Pilsen, Czechia, in 2017. In 2023,
he obtained the "Habilitation à Diriger des Recherches" (HDR) degree from the
University of Rennes, France.
He served as a Postdoctoral Fellow at imec/Ghent University, Belgium, until 2018,
followed by a position of a Scientist at the École Polytechnique Fédérale de Lausanne
(EPFL), Switzerland, until 2020. He then joined the French National Center for Scientific
Research (CNRS) as a Senior Scientist (fr. chargé de recherche). Dr. Nikolayev's
research focuses on the theoretical foundations and engineering of wireless
bioelectronics.
Dr. Nikolayev’s work has been recognized with several awards, including the Best
Ph.D. Dissertation Award from the Fondation Rennes 1. He is a laureate of the ERC
Starting Grant (2024), the Eiffel Excellence Doctoral Fellowship and the recipient of the
ANR–NRF PRCI Research Fellowship (2021), jointly awarded by the Agence Nationale
de la Recherche, France, and the National Research Foundation, Singapore. Since
2022, Dr. Nikolayev has served as an elected member of the Section 8 of the French
National Committee for Scientific Research.
\end{IEEEbiography}

\begin{IEEEbiography}[{\includegraphics[width=1in,height=1.25in,clip,keepaspectratio]{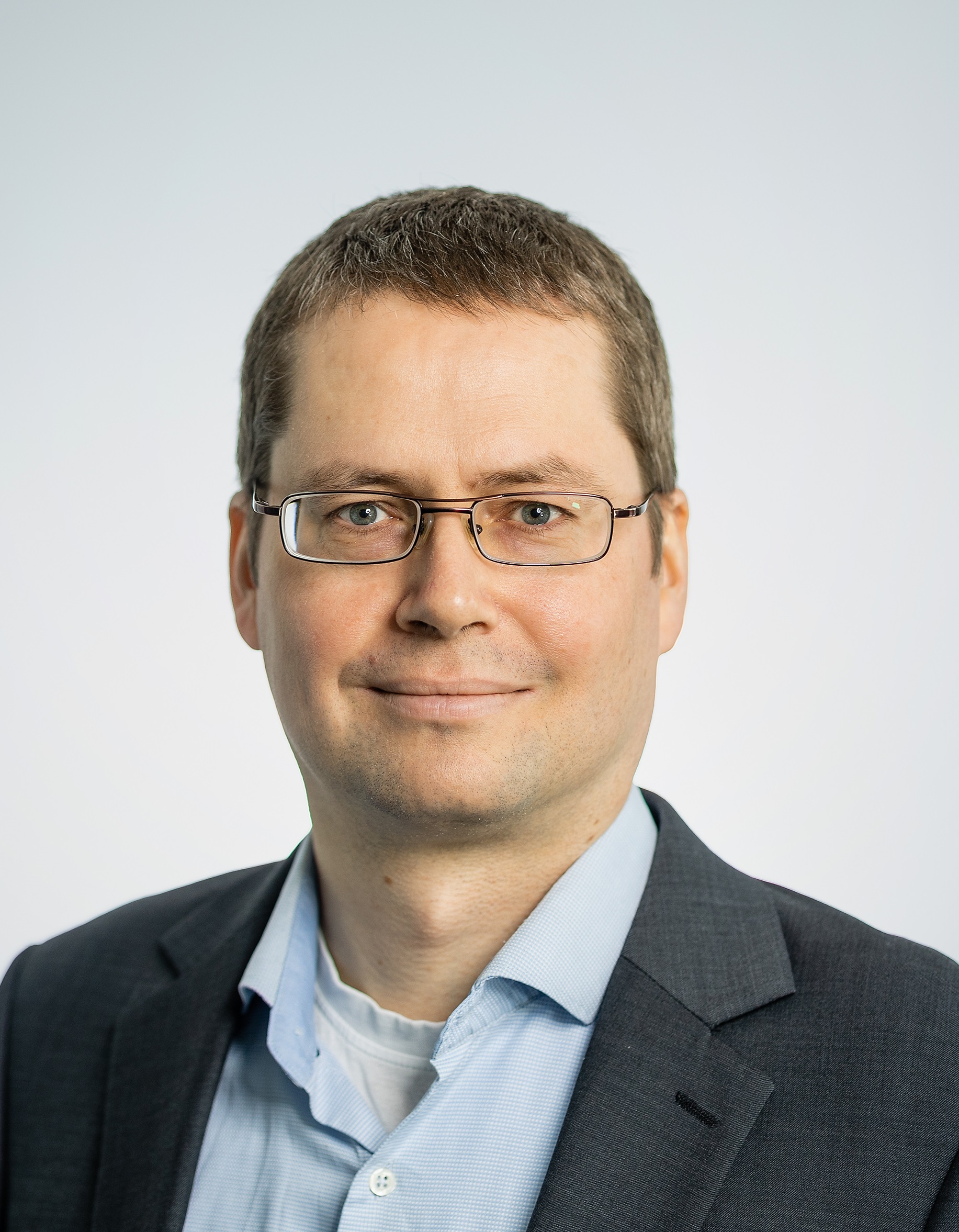}}]{P. Birkholz}
Peter Birkholz (Member, IEEE) was born in Rostock, Germany, in 1978. He received the Diploma in computer science and the Ph.D. degree (with distinction) in signal processing from the University of Rostock, Germany, in 2002 and 2005, respectively. He was a Research Associate with the University of Rostock from 2005 to 2009, and with the Department of Phoniatrics, Pedaudiology, and Communication Disorders, RWTH Aachen University, Aachen, Germany, from 2009 to 2014. He became a Junior Professor of Cognitive Systems with TU Dresden, Germany, in 2014, and a full Professor of Speech Technology and Cognitive Systems with the TU Dresden, Germany, in 2020. His research interests include speech production, articulatory speech synthesis, computational neuroscience, silent speech interfaces, and measurement techniques for speech research. Dr. Birkholz was the recipient of the Joachim-Jungius Prize in 2006 by the University of Rostock for his dissertation on articulatory speech synthesis, and of the Klaus-Tschira Award for Achievements in Public Understanding of Science in 2006.
\end{IEEEbiography}

\end{document}